\documentstyle[12pt]{article}

\def\hybrid{\topmargin -20pt	\oddsidemargin 0pt
	\headheight 0pt	\headsep 0pt
	\textwidth 6.25in	
	\textheight 9.5in	
	\marginparwidth .875in
	\parskip 5pt plus 1pt	\jot = 1.5ex}

\hybrid

\def\baselinestretch{1.2}

\catcode`\@=11

\def\marginnote#1{}
\newcount\hour
\newcount\minute
\newtoks\amorpm
\hour=\time\divide\hour by60
\minute=\time{\multiply\hour by60 \global\advance\minute by-\hour}
\edef\standardtime{{\ifnum\hour<12 \global\amorpm={am}%
	\else\global\amorpm={pm}\advance\hour by-12 \fi
	\ifnum\hour=0 \hour=12 \fi
	\number\hour:\ifnum\minute<10 0\fi\number\minute\the\amorpm}}
\edef\militarytime{\number\hour:\ifnum\minute<10 0\fi\number\minute}

\def\draftlabel#1{{\@bsphack\if@filesw {\let\thepage\relax
   \xdef\@gtempa{\write\@auxout{\string
      \newlabel{#1}{{\@currentlabel}{\thepage}}}}}\@gtempa
   \if@nobreak \ifvmode\nobreak\fi\fi\fi\@esphack}
	\gdef\@eqnlabel{#1}}
\def\@eqnlabel{}
\def\@vacuum{}
\def\draftmarginnote#1{\marginpar{\raggedright\scriptsize\tt#1}}

\def\draft{\oddsidemargin -.5truein
	\def\@oddfoot{\sl preliminary draft \hfil
	\rm\thepage\hfil\sl\today\quad\militarytime}
	\let\@evenfoot\@oddfoot	\overfullrule 3pt
	\let\label=\draftlabel
	\let\marginnote=\draftmarginnote
   \def\@eqnnum{(\theequation)\rlap{\kern\marginparsep\tt\@eqnlabel}%
\global\let\@eqnlabel\@vacuum}  }

\def\preprint{\twocolumn\sloppy\flushbottom\parindent 2em
	\leftmargini 2em\leftmarginv .5em\leftmarginvi .5em
	\oddsidemargin -.5in	\evensidemargin -.5in
	\columnsep .4in	\footheight 0pt
	\textwidth 10.in	\topmargin  -.4in
	\headheight 12pt \topskip .4in
	\textheight 6.9in \footskip 0pt
	\def\@oddhead{\thepage\hfil\addtocounter{page}{1}\thepage}
	\let\@evenhead\@oddhead	\def\@oddfoot{}	\def\@evenfoot{} }

\def\numberbysection{\@addtoreset{equation}{section}
	\def\theequation{\thesection.\arabic{equation}}}

\def\underline#1{\relax\ifmmode\@@underline#1\else
	$\@@underline{\hbox{#1}}$\relax\fi}

\def\titlepage{\@restonecolfalse\if@twocolumn\@restonecoltrue\onecolumn
     \else \newpage \fi \thispagestyle{empty}\c@page\z@
	\def\thefootnote{\fnsymbol{footnote}} }

\def\endtitlepage{\if@restonecol\twocolumn \else \newpage \fi
	\def\thefootnote{\arabic{footnote}}
	\setcounter{footnote}{0}}  

\catcode`@=12
\relax

\def\figcap{\section*{Figure Captions\markboth
	{FIGURECAPTIONS}{FIGURECAPTIONS}}\list
	{Figure \arabic{enumi}:\hfill}{\settowidth\labelwidth{Figure
999:}
	\leftmargin\labelwidth
	\advance\leftmargin\labelsep\usecounter{enumi}}}
 \relax
\def\tablecap{\section*{Table Captions\markboth
	{TABLECAPTIONS}{TABLECAPTIONS}}\list
	{Table \arabic{enumi}:\hfill}{\settowidth\labelwidth{Table
999:}
	\leftmargin\labelwidth
	\advance\leftmargin\labelsep\usecounter{enumi}}}
 \relax
\def\reflist{\section*{References\markboth
	{REFLIST}{REFLIST}}\list
	{[\arabic{enumi}]\hfill}{\settowidth\labelwidth{[999]}
	\leftmargin\labelwidth
	\advance\leftmargin\labelsep\usecounter{enumi}}}
 \relax
\makeatletter
\newcounter{pubctr}
\def\publist{\@ifnextchar[{\@publist}{\@@publist}}
\def\@publist[#1]{\list
	{[\arabic{pubctr}]\hfill}{\settowidth\labelwidth{[999]}
	\leftmargin\labelwidth
	\advance\leftmargin\labelsep
	\@nmbrlisttrue\def\@listctr{pubctr}
	\setcounter{pubctr}{#1}\addtocounter{pubctr}{-1}}}
\def\@@publist{\list
	{[\arabic{pubctr}]\hfill}{\settowidth\labelwidth{[999]}
	\leftmargin\labelwidth
	\advance\leftmargin\labelsep
	\@nmbrlisttrue\def\@listctr{pubctr}}}
 \relax
\makeatother
\newskip\humongous \humongous=0pt plus 1000pt minus 1000pt

\newif\ifdtup

\relax

\def\be{\begin{equation}}
\def\ee{\end{equation}}
\def\ba{\begin{eqnarray}}
\def\ea{\end{eqnarray}}

\def\del{\partial}

\def\r{\rho}
\def\a{\alpha}

\def\b{\beta}

\def\g{\gamma}
\def\G{\Gamma}
\def\d{\delta}

\def\e{\epsilon}

\def\th{\theta}

\def\m{\mu}
\def\n{\nu}
\def\om{\omega}
\def\Om{\Omega}
\def\l{\lambda}
\def\L{\Lambda}
\def\s{\sigma}

\def\bs{\bigskip}
\def\no{\noindent}

\def\qq{\qquad}

\def\IR{\relax{\rm I\kern-.18em R}}

\def \ha {{1\over 2}}

\def \ov {\over}

\def\IR{\relax{\rm I\kern-.18em R}}

\def\IR{\relax{\rm I\kern-.18em R}}

\begin{document}

\renewcommand{\theequation}{\thesection.\arabic{equation}}
\newcommand{\beq}{\begin{equation}}
\newcommand{\eeq}[1]{\label{#1}\end{equation}}
\newcommand{\ber}{\begin{eqnarray}}
\newcommand{\eer}[1]{\label{#1}\end{eqnarray}}
\begin{titlepage}
\begin{center}

\hfill CERN--TH/96--89\\
\hfill THU--96/19\\
\hfill hep--th/9604195\\

\vskip .4in

{\large \bf UNIVERSAL ASPECTS OF STRING PROPAGATION\\ }
{\large \bf ON CURVED BACKGROUNDS }

\vskip 0.5in

{\bf Ioannis Bakas}
\footnote{e--mail address: BAKAS@SURYA11.CERN.CH }
\vskip .1in

{\em Theory Division, CERN, 1211 Geneva 23, Switzerland, and\\
Department of Physics, University of
Patras,
26110 Patras, Greece}
\footnote{Permanent address} 

\vskip .3in

{\bf Konstadinos Sfetsos}
\footnote{e--mail address: SFETSOS@FYS.RUU.NL}\\
\vskip .1in

{\em Institute for Theoretical Physics, Utrecht University\\
     Princetonplein 5, TA 3508, The Netherlands}\\

\vskip .2in

\end{center}

\vskip .6in

\begin{center} {\bf ABSTRACT } \end{center}
\begin{quotation}\noindent

\no
String propagation on $D$--dimensional 
curved backgrounds with Lorentzian signature is formulated 
as a geometrical problem of embedding surfaces. 
When the spatial part of the background corresponds to 
a general WZW model for a compact group, 
the classical dynamics of the physical degrees of freedom is 
governed by the coset conformal
field theory $SO(D-1)/SO(D-2)$, which is universal 
irrespective of the particular WZW model.
The same holds for string propagation on 
$D$--dimensional flat space. The integration of the corresponding 
Gauss--Codazzi equations requires the introduction of 
(non--Abelian) parafermions in differential geometry.

\vskip .2in

\noindent

\end{quotation}
\vskip .2cm
April 1996\\
\end{titlepage}

\def\baselinestretch{1.2}
\baselineskip 16 pt
\noindent

\section{ Introduction }
\setcounter{equation}{0}

String propagation on a given background defines an embedding
problem in differential geometry. Choosing, whenever possible,
the temporal gauge one may solve the Virasoro constraints
and consider the non--linear dynamics governing the physical 
degrees of freedom of the string. Simple counting shows that for 
$D$--dim backgrounds the physical degrees of freedom satisfy a 
coupled system of $D-2$ differential equations, 
which are defined on the 2--dim string world--sheet and they are 
non--linear due to the quadratic form of the Virasoro constraints.
Our primary aim is to investigate the integrability of these 
equations and explore some of their universal aspects for a 
wide class of backgrounds.

Lund and Regge considered this problem several years ago for
string propagation on flat 4--dim Minkowski space,
in the presence of a Kalb--Ramond field as well \cite{LuRe}.
This geometrical approach was subsequently generalized to
$D\geq 5$ \cite{barba}.
It became clear more recently \cite{Baso3}
that the dynamics of the physical 
degrees of freedom in the $D=4$ case admits a Lagrangian 
formulation as an $SO(3)/SO(2)$ gauged WZW model.
However, for $D\geq 5$ an analogous Lagrangian description
using cosets to model the dynamics of the $D-2$
physical degrees of freedom has been lacking.
The technical problem that arises here is 
finding the appropriate non--local field variables to integrate the
underlying Gauss--Codazzi equations of the embedding.
We solve this problem by introducing, just from
purely geometrical considerations, the non--Abelian parafermions
of the coset model $SO(D-1)/SO(D-2)$ and show that the
chiral equations they obey \cite{BSthree} are equivalent to 
the Gauss--Codazzi embedding equations.
Hence, string dynamics on $D$--dim flat Minkowski space,
after we solve the Virasoro constraints, 
is governed by the semi--classical
geometry of the conformal field theory 
coset $SO(D-1)/SO(D-2)$ \cite{BSthree,BShet}.

An interesting generalization of this program includes Lorentzian
backgrounds of the product form
$R\otimes K_{D-1}$, where $K_{D-1}$ is a WZW model
for a semi--simple compact group. 
The integration of the Gauss--Codazzi 
equations for these backgrounds is similar to flat space in that
$SO(D-1)/SO(D-2)$ parafermions are also used, thus exhibiting
a universal behavior irrespectively of the particular WZW model
$K_{D-1}$. The coset space structure of the physical degrees of
freedom of the free string is rather remarkable, leading to the 
world--sheet integrability of the underlying non--linear equations.
Using the parafermion variables of the Gauss--Codazzi equations
one may easily find chiral $W_{\infty}$ symmetries as hidden 
on--shell symmetries of the classical theory. Our results shed
new light into the differential geometry of embedding surfaces using
concepts and field variables, which so far have been natural 
only in conformal field theory. 

The organization of this paper is as follows: In section 2 we set up the
Gauss--Codazzi equations for string propagation on $D$--dim curved
space and determine a wide class of backgrounds that allow for their
integration.  We expose the universal aspects of string dynamics
for Lorentzian backgrounds whose spatial part is either flat space
or a WZW model based on a general compact group.
In section 3 we use the $SO(D-1)/SO(D-2)$ WZW model to describe 
systematically the dynamics of the physical degrees of freedom 
and present  explicit results for $D=4$ and $D=5$.
Finally, in section 4 we comment on various other generalizations
and the quantization of strings before or after solving the classical
Virasoro constraints.
Connections with reduced $\s$--models \cite{Pohlalloi}--\cite{bapa} and 
the associated systems of symmetric space sine--Gordon models are 
also discussed.

\section{String dynamics and embedding surfaces}
\setcounter{equation}{0}
We first review 
relevant parts from the theory of embedding surfaces 
in the context of Riemannian geometry 
(see for instance \cite{Eisenhart}). Then we consider classical
string propagation on backgrounds with Lorentzian signature
and we formulate the problem of determining the dynamics of the
physical modes as a geometrical problem of surface embedding,
after solving the Virasoro constraints in the temporal gauge. 
At the end we specialize to backgrounds with spatial 
part corresponding to flat space or WZW models based on general
semi--simple compact groups.

\subsection*{ Gauss--Codazzi equations: Generalities}

Consider a $D$--dim space $M_D$ with 
line element ($\equiv$
fundamental quadratic form) given by
\be 
ds^2_D = G_{\m\n}(y) dy^\m dy^\n ~ , ~~~~~
\m,\n = 1,\dots , D ~ .
\label{dsD}
\ee
A $d$--dim subspace $M_d$ of $M_D$ with local
coordinates $x^i$, $i=1,\dots , d$ 
may be considered as an embedded surface with defining equations 
$y^\m = y^\m (x^1,\dots , x^d)$.
The line element in $M_d$ will be denoted by
\be
 ds^2_d = g_{ij}(x) dx^i dx^j ~ , ~~~~~
i,j = 1,\dots , d ~ .
\label{dsd}
\ee
The restriction of (\ref{dsD}) in $M_d$ should be equivalent to 
(\ref{dsd}). Thus we have the relation\footnote{ We will
use the notation
$y^\m_{,i}\equiv {\del y^\m\ov \del x^i}$. Covariant derivatives
on $M_D$ and on $M_d$ will be denoted by $D_\m$ and
$D_i$ respectively. The $y^\m(x)$'s are scalars
with respect to covariant differentiation on $M_d$, 
i.e., $D_i y^\m = y^\m_{,i}$.}
\be
g_{ij}(x) = G_{\m\n}(y) y^\m_{,i} y^\n_{,j} ~ .
\label{gij}
\ee
The embedded surface is completely specified by 
the set of vectors $\{\xi^\m_\s,~ \s=d+1,\dots , D\}$
normal to it. These are chosen to satisfy the orthonormalization
conditions
\be 
G_{\m\n} \xi^\m_\s \xi^\n_\tau = \d_{\s\tau} ~ ,
\label{gxx}
\ee
and by definition are also orthogonal to the tangent vectors
to the surface $y^\m_{,i}$:
\be
G_{\m\n} y^\m_{,i} \xi^\n_\s = 0 ~ .
\label{gyx}
\ee
The set of vectors $\{y^\m_{,i}, \xi^\m_\s\}$ satisfy 
the completeness relation in $M_D$:
\be
g^{ij} y^\m_{,i} y^\n_{,j} + 
\xi^\m_\s \xi^\n_\tau \d^{\s\tau} = G^{\m\n} ~ .
\label{commD}
\ee

The dynamics of the embedded surface is determined from 
the evolution of the vectors $y^\m_{,i}$ and $\xi^\m_\s$ 
as functions of the variables $x^i$ in $M_d$.
The corresponding equations are determined by repeated 
covariant differentiations of 
(\ref{gij})--(\ref{gyx}) and subsequent 
algebraic manipulations. Here we will only present the 
result leaving out the detailed proofs which can be found in 
\cite{Eisenhart}. We recall 
the concept of the second fundamental quadratic form with components
defined as
\be
\Om^\s_{ij} = G_{\m\n} \xi^\m_\s \Big( 
D_iD_j y^\n + \G^\n_{\l\a} y^\l_{,i} y^\a_{,j} \Big) ~ .
\label{omsij}
\ee
It is obvious that it is a symmetric 
tensor in $M_d$, i.e., $\Om^\s_{ij}=\Om^\s_{ji}$.
We also define the torsion ($\equiv$ third fundamental form)
in $M_d$
\be
\m^{\s\tau}_i = 
G_{\m\n} \xi^\m_\s \Big( \xi^\n_{\tau,i} + 
\G^\n_{\l\a} \xi^\l_\tau y^\a_{,i} \Big) ~ .
\label{mist}
\ee
Though not immediately obvious it can be shown that 
it is antisymmetric,
i.e., $\m_i^{\s\tau}+\m_i^{\tau\s}=0$. With the above
definitions the evolution equations can be written as
\be
D_iD_j y^\m = \Om^\s_{ij} \xi^\m_\s  
-  \G^\m_{\n\l} y^\n_{,i} y^\l_{,j} ~ ,
\label{dijy}
\ee
and
\be
\xi^\m_{\s,i}= - \Om^\s_{ij} g^{jk} y^\m_{,k} + 
\mu_i^{\tau\s} \xi^\m_\tau
- \G^\m_{\l\a} y^\l_{,i} \xi^\a_\s ~ .
\label{xmi}
\ee
The careful reader will notice that for curves ($d=1$) 
in 3--dim Euclidean
space, the equations (\ref{dijy}) and (\ref{xmi}) 
reduce to the well known Serret--Frenet formulae.

It is a quite straightforward but tedious procedure 
to derive the necessary conditions for the existence of solutions
to (\ref{dijy}) and (\ref{xmi}). The resulting 
compatibility equations are given by
\ba
R_{ijkl} & = &
R_{\m\n\a\b} y^\m_{,i} y^\n_{,j} y^\a_{,k} y^\b_{,l}
+ \Om^\tau_{k[i} \Om^\tau_{j]l} ~ ,
\label{rijkl} \\
D_{[k} \Om^\s_{j]i} & = & \m_{[k}^{\tau\s} \Om^\tau_{j]i}
+ R_{\m\n\a\b} y^\m_{,i} y^\a_{,j} y^\b_{,k} \xi^\n_\s ~ ,
\label{Dkijs} 
\ea
and
\be
D_{[k} \m_{j]}^{\s\tau} + \m_{[j}^{\r\s} \m_{k]}^{\r\tau}
+ \Om^\s_{l[j} \Om^\tau_{k]i} g^{li} 
+ R_{\m\n\a\b} y^\m_{,j} y^\n_{,k} \xi^\a_\s \xi^\b_\tau = 0 ~ .
\label{Dkmts}
\ee
Equations (\ref{rijkl}) and (\ref{Dkijs}) 
for the case of a 2--dim surface
in 3--dim Euclidean space are known as the Gauss--Codazzi 
equations, whereas (\ref{Dkmts})
for the case of a surface immersed in Euclidean space 
is known as the Ricci equation.
In general, the number of unknown functions in the embedding equations
of a space $M_d$ in $M_D$ 
exceeds the number of equations.
However, the extra functions may be eliminated using
the freedom to perform local transformations in the normal space to
the surface that 
rotate $\Om^\s_i$ and $\m^{\s\tau}_i$, also using
any additional information that might be in our disposal.
The precise mechanism, for the cases
of interest in this paper, will be
considered in detail in the next subsection.

\subsection*{String evolution in $M_D= R \otimes K_{D-1}$ }
 
We consider classical propagation of closed strings on a 
$D$--dim background that is 
the direct product of the real line $R$ (contributing a minus 
in the signature matrix)
and a general manifold (with Euclidean signature) $K_{D-1}$, 
i.e., $M_D= R \otimes K_{D-1}$.
The corresponding target space variables are 
$y^0(\s^+,\s^-)$ and 
$y^\m(\s^+,\s^-)$ with $\m=1,\dots , D-1$.
Here $\s^\pm= \ha(\tau\pm \s)$, where
$\tau$ and $\s$ are the natural time and spatial variables 
on the world--sheet $\Sigma$.
Then,
the 2--dim $\s$--model action is given by
\be
S= \ha \int_\Sigma Q^+_{\m\n} \del_+ y^\m \del_- y^\n 
- \del_+ y^0 \del_- y^0 ~ , ~~~~~ 
Q^+_{\m\n} = G_{\m\n} + B_{\m\n} ~ ,
\label{smoac}
\ee
where $G$, $B$ are the metric 
and antisymmetric tensor fields
corresponding to the non--trivial part 
of the string background.
The classical equations of motion are given by
\ba
\d y^0 : && ~~~~~ \del_+ \del_- y^0 = 0 ~ ,
\label{dy0} \\
\d y^\m : && ~~~~~ \del_+ \del_- y^\m 
+ (\G^-)^\m_{\n\l} \del_+ y^\n \del_- y^\l = 0 ~  ,
\label{dym}
\ea
where $(\G^\pm)^\m_{\n\l}=\G^\m_{\n\l} \pm\ha H^\m_{\n\l}$
are the generalized connections that include the string torsion
$H_{\m\n\l}\equiv \del_{[\m} B_{\n\l]}$.
We have implicitly imposed the conformal 
gauge in writing (\ref{smoac}).
Hence, the classical equations of motion
are supplied with the constraints 
\be
T_{\pm\pm}\equiv {1\ov 4} G_{\m\n} \del_\pm y^\m \del_\pm y^\n 
- {1\ov 4} \del_\pm y^0 \del_\pm y^0 = 0~ .
\label{Tpmpm}
\ee

The conformal gauge allows for transformations 
$\s^\pm\to f^\pm(\s^\pm)$,
which can be used in a way consistent with the equations of motion 
(\ref{dy0}), (\ref{dym}). We choose the so called temporal 
gauge, where $y^0=\tau$. Then (\ref{dy0}) is 
trivially satisfied whereas (\ref{dym}) remains unaffected 
since $G$ and $B$ are independent of $y^0$.
The constraints (\ref{Tpmpm}) take the form
\be
G_{\m\n} \del_\pm y^\m \del_\pm y^\n = 1 ~ .
\label{cooss}
\ee
For later use we define an angular variable $\th$
via the relation
\be
G_{\m\n} \del_+ y^\m \del_- y^\n = \cos \th ~ .
\label{angu}
\ee
The Euclidean signature of $K_{D-1}$ warrants the reality of $\th$.

Clearly in the temporal gauge we may restrict our analysis
entirely on $K_{D-1}$ and on 
the projection of the string world--sheet $\Sigma$ on the
$y^0=\tau$ hyperplane, following the spirit of the Lund--Regge 
analysis \cite{LuRe}. The resulting 2--dim surface
$S$ has Euclidean signature with metric given by
\ba
ds^2 & =&  G_{\m\n} dy^\m dy^\n \nonumber \\
& = & G_{\m\n} \Big( \del_+ y^\m \del_+ y^\n d{\s^+}^2
+ \del_- y^\m \del_- y^\n d{\s^-}^2 
+ 2 \del_+ y^\m \del_- y^\n d\s^+ d\s^- \Big )~ .
\label{dsS}
\ea
Using the constraints (\ref{cooss}) and the 
definition (\ref{angu}) we obtain from (\ref{dsS}) the
expression
\be
ds^2 = d{\s^+}^2 + d{\s^-}^2 + 2 \cos\th d\s^+ d\s^- ~ .
\label{dsS2}
\ee
Thus, for $y^0=\tau$, determining the classical
evolution of the string is equivalent to 
the problem of embedding the 2--dim surface $S$
with metric (\ref{dsS2}) on the $(D-1)$--dim space
$K_{D-1}$. Hence, the general analysis we have presented in the
previous subsection becomes relevant to string theory 
at this point.

For further convenience we present the 
expressions for the non--vanishing Christoffel symbols and
the Riemann curvature of the metric
(\ref{dsS2}):
\be 
\G^{\pm}_{\pm\pm} = \cot \th \del_\pm \th ~ , ~~~~
\G^\pm_{\mp\mp} = -{1\ov \sin\th} \del_\mp \th ~ , ~~~~
R_{+-+-}= - \sin\th \del_+ \del_- \th ~ .
\label{chrrie}
\ee
Contracting (\ref{dym}) with $G_{\m\a} \xi^\a_{\s}$, where
$\s=3,\dots , D-1$, and using (\ref{gxx}) we obtain
\be
\Om^\s_{+-} = \Om^\s_{-+}= \ha H_{\m\n\l} \xi^\m_{\s}
\del_+ y^\n \del_- y^\l ~, ~~~~~ \s=3,\dots , D-1 ~ .
\label{oms}
\ee
Contracting with $G_{\m\a} \del_\pm y^\a$ 
and using (\ref{cooss}) we obtain instead an identity and thus
have no additional restrictions. Hence, 
the information contained in the $D-1$ classical 
equations (\ref{dym}) is entirely encoded in the components
of the second fundamental form (\ref{oms}) and in the two
constraints (\ref{cooss}). It will be convenient to modify
the torsion $\m_\pm^{\s\tau}$ defined by (\ref{mist}),
using a term that includes the string torsion for $i=\pm$:
\ba
M_\pm^{\s\tau} & \equiv &  \m_\pm^{\s\tau} \pm \ha H_{\m\n\l}
\xi^\m_\s \xi^\n_\tau \del_\pm y^\l \nonumber \\
& = & G_{\m\n} \xi^\m_\s \Big( \del_\pm \xi^\n_\tau + 
(\G^\pm)^\n_{\l\a} \xi^\l_\tau \del_\pm y^\a \Big) ~ .
\label{Mnew}
\ea
It is evident that, similarly to $\m^{\s\tau}_\pm$, 
$M^{\s\tau}_\pm$ is also antisymmetric, and thus non--trivial
only for target spaces with dimension $D\geq 5$.
After some tedious algebraic manipulations,
equations (\ref{rijkl})--(\ref{Dkmts}) for the remaining components
of the second fundamental form $\Om^\s_{\pm\pm}$ and
for the modified torsion $M_\pm^{\s\tau}$
can be cast into the following form:
\ba
\Om^\tau_{++} \Om^\tau_{--} + \sin\th \del_+ \del_- \th
& = & - R^+_{\m\n\a\b} 
\del_+ y^\m \del_+ y^\a \del_- y^\n \del_- y^\b ~ ,
\label{gc1} \\
\del_{\mp} \Om^\s_{\pm\pm} - M_\mp^{\tau\s} \Om^\tau_{\pm\pm} 
-{1\ov \sin\th} \del_\pm\th \Om^\s_{\mp\mp}
& = & R^\mp_{\m\n\a\b} 
\del_\pm y^\m \del_\pm y^\a \del_\mp y^\b \xi^\n_\s ~ ,
\label{gc2}
\ea
and
\be
\del_+ M_-^{\s\tau} - \del_- M_+^{\s\tau} 
- M_-^{\r[\s} M_+^{\tau]\r} 
+ {\cos\th \ov \sin^2\th} \Om^{[\s}_{++} \Om^{\tau]}_{--} 
= \big(R^-_{\m\n\a\b}- D^-_\m H_{\n\a\b} \big)
\del_+ y^\m \del_- y^\n \xi^\a_\s \xi^\b_\tau ~ ,
\label{gc3}
\ee
where the curvatures are defined using the generalized 
connections $(\G^\pm)^\m_{\n\l}$,
\be
R^\pm_{\m\n\a}{}^\b = - \del_{[\m} (\G^\pm)^\b_{\n]\a}
+ (\G^\mp)^\g_{\a [\m} (\G^\pm)^\b_{\n]\g} ~ ,
\label{curva}
\ee
and similarly for the covariant derivatives $D^-_\m$ and 
$D^+_\m$.

Next, counting the number of the embedding equations in 
(\ref{gc1})--(\ref{gc3}) we find that there are 
$1+ 2(D-3) +\ha (D-3)(D-4)$ of them,
whereas the number of the unknown functions
$\th$, $\Om^\s_{\pm\pm}$ and $M_\pm^{\s\tau}$ is
$1+ 2 (D-3) + (D-3) (D-4)$. Hence, for $D\geq 5$ there are
$\ha (D-3)(D-4)$ more unknown functions than equations.
Notice, however, that the system (\ref{gc1})--(\ref{gc3}) is
invariant under local transformations
on the world--sheet generated by
\be
\xi^\m \to \L^{-1} \xi^\m ~ , ~~~~~ 
\Om_{\pm\pm}\to \L^{-1} \Om_{\pm\pm} ~ , ~~~~~ 
M_\pm \to \L^{-1} ( M_\pm + \del_\pm ) \L ~ ,
\label{gauge}
\ee
where $\L = \L(\s^+,\s^-)$ is an orthogonal matrix of $SO(D-3)$.
This gauge invariance
accounts for the extra (gauge) degrees of freedom 
in (\ref{gc1})--(\ref{gc3}) and can be used to eliminate them
(gauge fix).

\subsection*{ WZW backgrounds $K_{D-1}$ }

It seems an enormous task to make further progress with 
the embedding system of equations (\ref{gc1})--(\ref{gc3}) 
as it stands in all generality. There are two major difficulties.
First, the presence of source--like terms
depending explicitly on $\del_\pm y^\m$ and $\xi^\m_\s$
seems to prohibit us from integrating them, 
even partially. Second,
a Lagrangian description from which 
(\ref{gc1})--(\ref{gc3}) can be derived as equations of
motion is also lacking.

It is rather remarkable that both problems can be solved by 
considering for $K_{D-1}$ either flat space or
any WZW model based on a 
semi--simple compact group $G$, with $\dim(G)=D-1$. 
This is due to the identities \cite{zachos} 
\be
R^\pm_{\m\n\a\b} = D^\pm_\m H_{\n\a\b} = 0 ~ ,
\label{rdho}
\ee
which are  generally valid for any WZW model.
Then we completely get rid of the bothersome terms on the right
hand side of (\ref{gc1})--(\ref{gc3}).\footnote{Actually,
the same result
is obtained by demanding the weaker conditions 
$R^-_{\m\n\a\b}-D^-_\m H_{\n\a\b}=0$ and using the general
identity $R^-_{\m\n\a\b}-D^-_\m H_{\n\a\b}=
R^+_{\m\n\a\b}-D^+_\n H_{\m\a\b}$ and the property 
$R^+_{\m\n\a\b}=R^-_{\a\b\m\n}$. It would be interesting 
to find explicit examples where these weaker conditions hold.}
In order to show that a Lagrangian description exists, we first
extend the range of definition of 
$\Om^\s_{++}$ and $M_\pm^{\s\tau}$ by appending new components
defined as:
\be
\Om^2_{++}= \del_+ \th~ , ~~~~ 
M_+^{\s 2}= \cot \th \Om^\s_{++} ~ , ~~~~ 
M_-^{\s2} = -{1\ov \sin\th} \Om^\s_{--} ~ .
\label{exxtt}
\ee
Then equations (\ref{gc1})--(\ref{gc3}) can be recast into the
suggestive form 
\ba
&& \del_- \Om^a_{++} + M_-^{ab} \Om^b_{++} = 0 ~ , 
\label{new1} \\
&& \del_+ M_-^{ab} - \del_- M_+^{ab} + [M_+,M_-]^{ab} = 0 ~ ,
\label{new2}
\ea
where the new index $a=(2,\s)$. 
Notice that if we treat $\Om^a_{++}$ 
not as a row of the bigger matrix $M_+^{ab}$,
as suggested by (\ref{exxtt}), but as an independent vector, 
then the number of unknown
functions in (\ref{new1}) and (\ref{new2}) 
is augmented by $D-3$ compared to
the same number of functions in
equations (\ref{gc1})--(\ref{gc3}). 
However, there is a simultaneous 
enlargement of the local gauge symmetry 
from $SO(D-3)$ to $SO(D-2)$ that takes care of it. 
Such a gauge symmetry 
enlargement
can only be achieved if the underlying change of variables
is non--local. This will become more clear soon after the 
introduction of parafermions in the next section.

Equation (\ref{new2}) is a 
zero curvature condition for the matrices $M_\pm$ and it is 
solved (without worrying here about global issues related to the 
world--sheet topology) by $M_\pm = \L^{-1} \del_\pm \L$, 
where $\L \in SO(D-2)$. Then (\ref{new1}) can be written as
\be
\del_- (\L^{ab} \Om^b_{++}) = \del_- ( \L^{a2} \del_+ \th
+ \del_+ \L^{a2} \tan\th) = 0 ~ .
\label{new3}
\ee
The vector $\L^{a2}$ has unit length, i.e., $\L^{a2}\L^{a2}=1$.
We can incorporate this constraint by defining 
$Y^a=\L^{a2} \sin \th$. Then (\ref{new3}) assumes the form
\be
\del_- \left( {\del_+ Y^a \ov \sqrt{1-Y^2}} \right) = 0~ ,
~~~~~ Y^2\equiv Y^b Y^b~ , ~~~~~ a,b = 2,3,\dots ,D-1 ~ .
\label{fiin}
\ee
These equations were derived before in \cite{barba}, while 
describing
the dynamics of a free string propagating in $D$--dimensional
{\it flat} space--time. It is remarkable that these
equations remain 
unchanged even if the flat $(D-1)$--dim space--like part is replaced
by a curved background corresponding to a general WZW model.
In retrospect, we may attribute this unexpected result to the fact 
that a group space is parallelizable, and it 
can be made flat in the sense of (\ref{rdho}) 
with the addition of the appropriate amount of torsion.
It should be emphasized that although the compatibility 
equations are universal, 
the actual evolution equations of the normal and tangent
vectors to the surface 
are given by specializing (\ref{dijy}) and (\ref{xmi}) 
to $K_{D-1}$; they are certainly different from those 
of the flat space free string.

As we have already mentioned, it would be advantageous if 
(\ref{fiin}) (or an equivalent system) could be derived 
as classical equations of motion.
The key that will 
enable us next to construct the corresponding Lagrangian is 
the observation that (\ref{fiin}) 
imply chiral conservation laws, which 
are reminiscent of the
equations obeyed by classical
parafermions in coset models
\cite{BCR}.
In fact (\ref{fiin}) were derived as classical string equations
for gauged WZW models corresponding 
to $SO(D-1)/SO(D-2)$ cosets
in \cite{BSthree}; they are 
analytic continuations of the 
models $SO(D-3,2)/SO(D-3,1)$ that give rise to string propagation
in backgrounds with Lorentzian signature \cite{BN}.
We mention for completeness that they also arise 
in the massless limit of the
$SO(D)/SO(D-1)$ symmetric space sine--Gordon models, 
which were recently formulated as
integrable perturbations of the $SO(D-1)/SO(D-2)$ 
gauged WZW models \cite{bapa}.
Since (\ref{fiin}) themselves do not correspond to 
a Lagrangian system of equations,
our strategy in the following will be to perform a non--local change 
of variables that maps them into Lagrangian form. 
This non--local change of variables is highly non--intuitive 
in differential geometry, and only 
the correspondence with parafermions makes it natural.

\section { Dynamics of physical degrees of freedom }
\setcounter{equation}{0}

In this section we first briefly discuss some general aspects
of gauged WZW models in connection with the associated coset 
conformal field theories. Then we restrict our attention 
to $SO(D-1)/SO(D-2)$
coset models and establish a relation between the chiral conservation
laws obeyed by the corresponding
parafermions and the embedding equations (\ref{fiin}).
At the end we present explicit results for 
$D=4$ and $D=5$.

\subsection*{Lagrangian description and parafermions}

Recall that the gauged WZW action is \cite{SCH,gwzwall}
\be
S=I_{wzw}(g) + {k\ov \pi} \int 
{\rm Tr}\Big( A_+ \del_- g g^{-1} -  g^{-1}\del_+ g A_-
+ A_+ g A_- g^{-1} - A_+ A_-\Big) ~ ,
\label{GWZW}
\ee
where $g\in G$ and
$A_\pm$ are gauge fields valued in the Lie algebra of a 
subgroup $H\subset G$.
The corresponding field strength is
$F_{+-}= \del_+ A_- - \del_- A_+ - [A_+,A_-]$.
We also split indices as $A=(a,\a)$,
where $a\in H$ and $\a\in G/H$.
Variation of (\ref{GWZW}) with respect to all
fields gives the classical equations of motion
\ba
\d A_+ & :& ~~~~ \qq D_- g g^{-1} \big |_H  =0 ~ , \label{dap}
\\
\d A_- & : & ~~~~ \qq g^{-1} D_+ g  \big |_H   =0 ~ , \label{dam}
\\
\d g & : & ~~~~ \qq D_-(g^{-1} D_+ g) + F_{+-}=0 ~ . \label{dg} 
\ea
Imposing (\ref{dam}) on (\ref{dg}) yields the zero curvature 
condition $F_{+-}=0$ on--shell, and 
\be
D_-(g^{-1} D_+ g)\big |_{G/H} =0 ~ . 
\label{dgco} 
\ee

There are two commuting copies of an affine algebra
corresponding to a WZW action for a group $G$,
one for each chiral sector \cite{Wwzw}. 
A remnant of this algebra 
is also present in the gauged WZW model. 
We parametrize the gauge fields as 
$A_\pm =(\del_\pm h_\pm) h_\pm^{-1}$,
where $h_\pm \in H$.
Thus, $h_\pm$ are given in terms of $A_\pm$
as
\be
h_+^{-1} = {\rm P} e^{- \int^{\s^+} A_+}~ , ~~~~~
h_-^{-1} = {\rm P} e^{- \int^{\s^-} A_-}~ , 
\label{hphm}
\ee
where ${\rm P}$ stands for path ordering.
Using the gauge invariant group element
\be
f=h_+^{-1} g h_+ \in G ~ ,
\label{gainv}
\ee
and
the on--shell zero curvature condition $F_{+-}=0$,
we write equation (\ref{dgco}) as 
\be
\del_- \Psi_+ = 0 ~ , ~~~~~~~~~ 
\Psi_+ = {i k\ov \pi} f^{-1} \del_+ f \in G/H ~ .
\label{paraf}
\ee
Thus, the coset valued matrix $\Psi_+$ is chirally conserved. 

In fact, $\Psi_+$ are nothing but the classical parafermions
\cite{BCR}.
Since they have Wilson lines attached to them 
(cf. (\ref{gainv}), (\ref{hphm})) they are non--local objects. 
This is also reflected in the
algebra they obey \cite{BCR} (we drop + as a subscript and denote
$\s^+$ by $x$ or $y$),
\ba
\{ \Psi_\a(x),\Psi_\b(y)\} & = & - {k\ov \pi} \d_{\a\b}
\d^\prime (x-y)
- f_{\a\b\g} \Psi_\g (y) \d(x-y)
\nonumber \\
& & - {\pi\ov 2k}
f_{c\a\g} f_{c\b\d}~ \e(x-y) \Psi_\g(x) \Psi_\d (y) ~ ,
\label{poipar}
\ea
where the antisymmetric step function 
$\e(x-y)$ equals $+1~ (-1)$ if $x>y$ ($x<y$). 
The last term in (\ref{poipar}) is responsible 
for their non--trivial
monodromy properties and unusual statistics. 
In addition, conformal transformations are generated by 
$T_{++}=-{\pi\ov 2 k} \Psi_\a\Psi_\a$.

The 2--dim $\s$--model having the above 
infinite dimensional
symmetry is obtained by first
choosing a unitary gauge by fixing $\dim(H)$ 
variables among the total number of $\dim(G)$ parameters
of the group element $g$. Hence, there are
$\dim(G/H)$ remaining variables, which will be denoted by $X^\m$. 
Then, we eliminate the gauge fields
in (\ref{GWZW}) using their equation of motion 
(\ref{dap}), (\ref{dam})
\ba
A_+^a & = &+ i \big(C^T - I)^{-1}_{ab} 
L^b_\m \del_+ X^\m ~ ,
\nonumber \\
A_-^a & = & -i \big(C - I \big)^{-1}_{ab} 
R^b_\m \del_- X^\m  ~ ,
\label{apam}
\ea 
where the appropriate short--hand definitions are
\be
L^a_\m = -i {\rm Tr}(t^a g^{-1} \del_\m g)~ ,~~~~~
R^a_\m = -i {\rm Tr}(t^a  \del_\m g g^{-1})~ , ~~~~~
C^{ab}= {\rm Tr}(t^a g t^b g^{-1}) ~ .
\label{LRC}
\ee
Finally, the $\s$-model action is given by
\be 
S= I_{wzw}(g) - {k\ov \pi} \int_\Sigma
R^a_\m \big(C^T - I \big)^{-1}_{ab} L^b_\n 
\del_+ X^\m  \del_- X^\n ~ .
\label{dualsmo}
\ee

\subsection*{  $SO(D-1)/SO(D-2)$ coset structure}

We specialize now to the $SO(D-1)/SO(D-2)$ gauged
WZW model and show that 
(\ref{fiin}) is equivalent to 
the parafermion equation (\ref{paraf}).
We will essentially follow the analysis of \cite{BSthree} 
adopted to our present purposes.

The group element 
$g\in SO(D-1)$ in the right coset decomposition 
can be written as $g=\tilde h t $, where
\be
\tilde h = \left( \begin{array} {cc}
1 & 0 \\
  &  \\
0 & h \in SO(D-2) \\
\end{array}
\right)
\label{H}
\ee
and $t\in SO(D-1)/SO(D-2)$ is parametrized by a
$(D-2)$--dim vector $\vec X$ as  
\be
t = \left( \begin{array} {cc}
b & X^j \\
  &  \\
- X^i  & \d_{ij} - {1\ov b+1} X^i X^j \\
\end{array}\right) ~ , ~~~~~~ 
b \equiv \sqrt{1-\vec X^2} ~ .
\label{t}
\ee
The range of the parameters $X^i$ is restricted by $\vec X^2\leq 1$
and the value of $b$ is such that the matrix $t$ is an 
element of $SO(D-1)$ obeying $t^{-1}= t^T$.
Then we compute 
\ba
dt t^{-1}& =&  \left( \begin{array} {cc}
0 &  dX^j + {\vec X \cdot d\vec X \ov b(b+1)} X^j \\
 & \\
- dX^i - {\vec X \cdot d\vec X\ov b(b+1)} X^i  
& {1\ov b+1} dX^{[i} X^{j]} \\
\end{array}
\right)~ ,\nonumber \\ 
t^{-1} dt & = & \left( \begin{array} {cc}
0 &  dX^j + {\vec X \cdot d\vec X\ov b(b+1)} X^j  \\
  & \\
- dX^i - {\vec X \cdot d\vec X\ov b(b+1)} X^i  
& - {1\ov b+1} dX^{[i} X^{j]} \\
\end{array}
\right)~ . 
\label{dtt}
\ea

To find an explicit expression for the parafermions in (\ref{paraf})
we first rewrite the constraint (\ref{dam}) 
as 
\be
(f^{-1}\del_+ f)_{ij} = (T^{-1} \del_+ T)_{ij} +
(T^{-1} H^{-1} \del_+ H T)_{ij}=0 ~ , 
\label{conba}
\ee
where $H=h_+^{-1} h h_+$ and
$T=h_+^{-1} t h_+$. The explicit form of $T$ is as in
({\ref{t}) with 
$X^i\to Y^i \equiv X^j (h_+)^{ji}$. Notice that
since $\vec Y^2=\vec X^2$, the $Y^i$ are gauge invariant.
Then we solve for 
\be
(H^{-1} \del_+ H)_{ij} = {1\ov b(b+1)} \del_+ Y^{[i} Y^{j]} ~ .
\label{hi}
\ee
The parafermion in (\ref{paraf}) is computed by explicitly 
writing out $\Psi^i \equiv  {ik\ov \pi} (f^{-1} \del_+ f)_{0i}$ 
and utilizing (\ref{hi}).
The final result is \cite{BSthree}
\ba
&& \Psi^i  = {ik\ov \pi}
{\del_+ Y^i\ov \sqrt{1-\vec Y^2}}  =  
  {ik\ov \pi} {1\ov \sqrt{1-\vec X^2}} (D_+X)^j  h_+^{ji} ~ ,
\nonumber \\
&& (D_+X)^j =  \del_+ X^j - A_+^{jk} X^k ~ .
\label{equff}
\ea
Thus, the corresponding equation 
$\del_- \Psi^i = 0$ is precisely (\ref{fiin}).
The $Y^i$ are related to the
$\s$-model variables non--locally as
\be
Y^i = X^j (h_+)^{ji}~ , ~~~~~~~~
h_+^{-1}= {\rm P} e^{- \int^{\s^+} A_+}~ ,
\label{Yii}
\ee
where the gauge field $A_+$ is given by (\ref{apam}). This provides 
the necessary non--local change of variables that transform 
(\ref{fiin}) into a Lagrangian system of equations.

The representation matrices for $SO(D-1)$ are 
$(t_{AB})_{CD}=
\d_{C[A} \d_{B]D}$, where the indices split
as $A=(0,i)$ with $i=1,2,\dots , D-2$.
Then the algebra of the parafermions (\ref{poipar}) becomes
\be
\{ \Psi^i(x),\Psi^j(y)\} =  {k\ov 2 \pi} \d_{ij} \d^\prime (x-y) 
-{\pi\ov 2 k} \e(x-y) \Big( \d_{ij} \Psi(x)\cdot \Psi(y) 
- \Psi_j(x) \Psi_i(y) \Big) ~ .
\label{poi}
\ee
The absence of linear terms in $\Psi^i$ 
on the right hand side 
is due to the simple fact that $SO(D-1)/SO(D-2)$ 
is a symmetric space. 
Thus,
structure constants involving only coset space indices are zero.

It remains to choose a gauge and 
explicitly compute $A_+$ and the $(D-2)$--component
$\s$--model action (\ref{dualsmo}).
This has been done in another context
for $SO(3)/SO(2)$ (the only Abelian case) in \cite{BCR,WITbh},
for $SO(4)/SO(3)$ in \cite{BSthree} and for $SO(5)/SO(4)$ in
\cite{BShet}. 
Here, for the time being we proceed with a unified 
treatment of all $SO(D-1)/SO(D-2)$ models.
It is convenient to distinguish between the cases of $D$ 
even or odd integers.

\vskip .1 cm

\underline{$D=2 N + 2= {\rm even}$}: We have enough gauge freedom to
cast the orthogonal matrix $h\in SO(2N)$ and
the vector $\vec X$ into the form
\be
h =  \pmatrix{
\cos 2\phi_1 & \sin 2\phi_1 & 0 & \cdots & 0 & 0\cr
-\sin 2\phi_1 & \cos 2\phi_1 & 0 & \cdots & 0 & 0\cr
0 & 0 & 0 & \cdots & 0 & 0 \cr
\vdots & \vdots &\vdots & & \vdots & \vdots \cr
0 & 0 & 0 & \cdots & \cos 2\phi_N & \sin 2\phi_N \cr
0 & 0 & 0 & \cdots & -\sin 2\phi_N & \cos 2\phi_N  \cr } ~ ,~~~
\vec X = \left( \begin{array} {c}
0 \\ X_2 \\ 0 \\ X_4 \\ \vdots \\ 0 \\ X_{2N} \end{array} \right)~ .
\label{hdixn}
\ee
The total number of independent variables in $h$ and $\vec X$ is
$2N=D-2$, as it should be.

\vskip .1 cm

\underline{$D=2 N + 3= {\rm odd}$}: In such cases 
the orthogonal matrix $h\in SO(2N+1)$ and 
the vector $\vec X$ can be gauge fixed into the 
form
\be
h =  \pmatrix{
\cos 2\phi_1 & \sin 2\phi_1 & 0 & \cdots & 0 & 0 & 0\cr
-\sin 2\phi_1 & \cos 2\phi_1 & 0 & \cdots & 0 & 0 & 0\cr
0 & 0 & 0 & \cdots & 0 & 0 & 0\cr
\vdots & \vdots &\vdots & & \vdots & \vdots & \vdots \cr
0 & 0 & 0 & \cdots & \cos 2\phi_N & \sin 2\phi_N & 0\cr
0 & 0 & 0 & \cdots & -\sin 2\phi_N & \cos 2\phi_N & 0 \cr
0 & 0 & 0 & \cdots & 0 & 0 & 1 \cr } ~ ,~~~
\vec X = \left( \begin{array} {c}
0 \\ X_2 \\ 0 \\ X_4 \\ \vdots \\ 0 \\ X_{2N} \\ X_{2N+1}
\end{array} \right) ~ .
\label{hdixn1}
\ee
Again the total number of the remaining independent variables 
is $2N +1=D-2$, as it should be. 

Using the above gauge fixing together with (\ref{dtt}) and
the Polyakov--Wiegman formula, we find that the WZW action 
(\ref{dualsmo}) contributes to the total line element 
\be
ds^2_{wzw}=  d\vec \phi^2
+ {1\ov 2(1+ b)} d\vec X^2 +  {1+ 2 b\ov 4 b^2(1+b)^2}
(\vec X \cdot d\vec X)^2 ~ ,
\label{sodwzw}
\ee
and has zero contribution to the total antisymmetric tensor.
The contribution of the second term of (\ref{dualsmo}) is more 
complicated and will not be presented here in all
generality; of course, its effect will be taken into account in the 
specific examples below.

\subsection*{ Examples }

We will work out all the technical details in two examples. The first one
is the Abelian coset $SO(3)/SO(2)$ \cite{BCR}. 
In terms of our original problem it arises after solving the 
Virasoro constraints for
strings propagating on 4--dim Minkowski space or on the 
direct product of the real line $R$ and the WZW model for 
$SU(2)$, which is the only 3--dim non--Abelian
group for which a WZW action exists.
The second example is the simplest 
non--Abelian coset based on $SO(4)/SO(3)$ and was considered in
\cite{BSthree}. In our context it
arises in string
propagation on 5--dim Minkowski space or on the direct 
product of the real line $R$ and the WZW model based on 
$SU(2)\otimes U(1)$.

\vskip .1 cm

\underline{$SO(3)/SO(2)$}: Using (\ref{hdixn}) with $X_2=\sin 2\th$, 
we find that the solution for the gauge fields is
\be
A_\pm = \pmatrix{ 0 & 1\cr -1 & 0 } 
(1\mp \cot^2\th) \del_\pm \phi ~ ,
\label{gsu2}
\ee
and that the corresponding background has metric \cite{BCR} 
\be 
ds^2 = d\th^2  + \cot^2\th d\phi^2 ~ .
\label{S1}
\ee 
Using (\ref{equff}), the corresponding Abelian parafermions 
$\Psi_\pm = \Psi_2 \pm i\Psi_1$ assume the familiar form 
\be
\Psi_\pm = (\del_+ \th \pm i \cot\th \del_+ \phi) 
e^{\mp i \phi \pm i  \int \cot^2\th \del_+ \phi } ~ ,
\label{pasu2}
\ee
up to an overall normalization. 

The emergence of the $SO(3)/SO(2)$ parafermions can also be seen 
directly from the original system of embedding equations
(\ref{gc1})--(\ref{gc3}). Since the indices $\s,\tau$ take only one
value, the torsion matrix is $\m_\pm=0$. Then
equation (\ref{gc3}) is trivially satisfied, whereas (\ref{gc1}) and
(\ref{gc2}) give (after setting
$\Om_{\pm\pm}=\cot {\th\ov 2}\del_\pm \phi$) the following 
two equations:
\ba
&& \del_+\left(\cot^2 {\th\ov 2} \del_- \phi \right) 
+ \del_-\left(\cot^2 {\th\ov 2} \del_+ \phi \right) = 0 ~ ,
\nonumber \\
&& \del_+\del_- \th  + {\cos{\th\ov 2} \ov 2 \sin^3{\th\ov 2} }
\del_+\phi \del_- \phi = 0~ .
\label{bh2d}
\ea
These are the classical equations of motion of
the $SO(3)/SO(2)$ coset 
with metric (\ref{S1}) (up to rescaling of
$\th,\phi$ by a factor of 2) having the parafermions (\ref{pasu2})
as natural chiral objects.
In the present geometrical context equations (\ref{bh2d}) were first 
derived in \cite{LuRe}, whereas in \cite{Baso3} it was 
subsequently realized 
that they admit the $SO(3)/SO(2)$ coset interpretation we have
just mentioned.
It should be pointed out that for $D\ge5$ a
Lagrangian description for the embedding equations 
(\ref{gc1})--(\ref{gc3}) cannot be possibly found in general
without first
making contact with parafermions, due to the fact that the torsion
matrix $\m_\pm$ (or $M_\pm$) is non--trivial.

\vskip .1 cm

\underline{$SO(4)/SO(3)$}: We parametrize
$X_2= \sin2\th \cos \om $ and $X_3 = \sin2\th \sin\om$ 
and use the basis of $SO(3)$ representation matrices
\be
t_{12}=\left( \begin{array} {ccc}
0& 1 & 0 \\
-1 & 0 & 0 \\
0 & 0 & 0 \end{array} \right)~ , ~~~
t_{13}=\left( \begin{array} {ccc}
0& 0 & 1 \\
0 & 0 & 0 \\
-1 & 0 & 0 \end{array} \right)~ , ~~~
t_{23}=\left( \begin{array} {ccc}
0& 0 & 0 \\
0 & 0 & 1 \\
0 & -1 & 0 \end{array} \right)~ .
\label{t123}
\ee
Using (\ref{hdixn1}) and the expansion for the gauge fields 
$A_\pm =  \sum_{i<j} A_\pm^{ij} t_{ij}$
we find the solution
\ba
A^{12}_+  & = & -\left( {\cos 2\th \ov \sin^2\th \cos^2\om }
+ \tan^2\om {\cos^2\th -\cos^2\phi \cos 2\th\ov 
\cos^2\th \sin^2 \phi} \right)
\del_+\phi - \cot\phi \tan\om \tan^2 \th 
\del_+\om ~ ,\nonumber \\ 
A^{13}_+ & = &  \tan\om
{\cos^2\th -\cos^2\phi \cos 2\th\ov 
\cos^2\th \sin^2 \phi} \del_+\phi
+ \cot\phi \tan^2 \th \del_+\om ~ ,
\label{expap} \\
A^{23}_+ & = & \cot\phi \tan \om {\cos 2\th\ov \cos^2\th} 
\del_+ \phi - \tan^2 \th \del_+\om ~ .
\nonumber 
\ea
It turns out that an analogous expression for $A_-^{ij}$ 
can be found from (\ref{expap}) by
writing all $\th$--dependence in terms of $\cos2 \th$ and 
replacing $\cos 2\th$ by $ 1/\cos 2 \th$.
Then, the 
background metric is \cite{BSthree}
\be
ds^2 = d\th^2 + \tan^2\th (d\om + \tan\om \cot \phi d\phi)^2 
+ {\cot^2\th \ov \cos^2\om} d\phi^2 ~ ,
\label{ds3}
\ee
and the antisymmetric tensor is zero.
The parafermions of the $SO(4)/SO(3)$ coset are non--Abelian 
given by (\ref{equff}) with 
covariant derivatives (omitting an overall factor of 2)
\ba 
{(D_+X)^1 \ov \sqrt{1-\vec X^2}} & = & 
{\cot\th\ov \cos\om} \del_+ \phi ~ ,
\nonumber \\
{ (D_+X)^2 \pm i (D_+X)^3 \ov \sqrt{1-\vec X^2}} & = & 
e^{\pm i\om} \Big(\pm i \tan\th ( \tan\om \cot \phi \del_+ \phi
+ \del_+ \om ) + \del_+ \th \Big) ~ .
\label{DDDX}
\ea
As a check, one may verify that $\Psi^i\Psi^i = {1\ov 1-\vec X^2} (D_+ X)^i
(D_+ X)^i$ is indeed proportional to the $T_{++}$--component
of the energy momentum tensor corresponding to a 
$\s$--model with metric (\ref{ds3}).

In addition to the two examples above, there also
exist explicit results for the coset $SO(5)/SO(4)$
\cite{BShet}. 
This would correspond in our context to string
propagation on a 6--dim Minkowski space or 
on the background 
$R$ times the $SU(2)\otimes U(1)^2$
WZW model.
It should be pointed out that there is no reason to demand 
conformal invariance for the backgrounds with
metrics (\ref{S1}) and (\ref{ds3}) because they arise in a 
different context describing the geometry of the 
physical degrees of freedom.


\section{ Conclusions }
\setcounter{equation}{0}

We have investigated some universal aspects of classical string dynamics
by integrating the Gauss--Codazzi equations of the corresponding
embedding problem.
We found for the class of $D$--dim backgrounds $R\otimes K_{D-1}$,
where $K_{D-1}$ is $R^{D-1}$ or the WZW model for a general $(D-1)$--dim
semi--simple compact group, that there are
$D-2$ physical degrees of freedom whose dynamics is governed by the
$SO(D-1)/SO(D-2)$ coset conformal field theory. The parafermion 
variables of this coset arise naturally in the present geometrical 
context, and so our results could be viewed as a 
link between conformal field
theory techniques and the classical differential geometry of 
embedding surfaces. 

There are two obvious extensions one can further make. First, suppose we 
start with a $D$--dim string background with signature ($2,D-2$). 
The ``spatial'' part of this background is now Lorentzian, and 
therefore one has to consider suitable analytic continuation of the
previous results. In particular, instead of the coset 
$SO(D-1)/SO(D-2)$ 
we find that the dynamics of the physical degrees of 
freedom is now given by the non--compact coset 
$SO(D-3,2)/SO(D-3,1)$. The simplest version of this for $D=4$
has already been considered in \cite{vega}.
Second, it is also interesting to consider various supersymmetric 
generalizations of the present framework. 

There are many similarities between classical string dynamics and the 
theory of ordinary 2--dim $\s$--models. The latter can also be viewed 
as describing the embedding of 2--dim surfaces 
into a group or coset space manifold, which in turn is embedded 
in flat space.
Exploiting classical conformal invariance, which is similar to choosing
the orthonormal gauge in string theory, amounts to reducing ordinary $\s$--models
to the so called symmetric space 
sine--Gordon models (SSG) \cite{Pohlalloi}--\cite{ARSI}. The
SSG models have been described as perturbations of conformal field
theory cosets \cite{bapa}; for example, 
the reduced $S^n=SO(n+1)/SO(n)$ $\s$--model
yields an integrable sine--Gordon perturbation 
of the $SO(n)/SO(n-1)$ coset conformal field theory. Hence, 
apart from the potential terms, and in the absence of string 
self--interactions, the structure of the kinetic terms is the same
for the two classes of embedding problems. It is interesting to note 
that other reduced $\s$--models for general symmetric spaces have been 
described using appropriately chosen gauged 
WZW cosets (plus perturbations).
Therefore, the parafermion variables of the
corresponding coset conformal field theories (at and away from the 
conformal point) also play a key role in the integration of the
embedding equations.

Finally, an interesting issue is the quantization of string theory. 
There are two different methods of quantizing constrained systems,
either by solving the classical constraints and then quantize
directly the physical degrees of freedom, or by quantizing 
the unconstrained degrees of freedom and then impose the constraints
as quantum conditions on the physical states. 
It is well known that in general these two 
methods of quantization are not equivalent, in particular when the
constraints have quadratic form as in string theory. 
Quantization of string theory usually proceeds using the second method,
but in the
present framework the physical degrees of freedom should be quantized
directly using the quantization of the associated parafermions.
Exploring this issue further is an interesting problem.


\bs\bs 

\centerline{\bf Acknowledgments }
\no
One of us (I.B.) wishes to thank the Institute for Theoretical Physics
in Utrecht for hospitality and financial support during the last stages
of this research.
The work of K.S. was carried out with the financial support 
of the European Union Research Program 
``Training and Mobility of Researchers'' and is part of the project
``String Duality Symmetries and Theories with 
Space--Time Supersymmetry'', under contract ERBFMBICT950362.


\end{document}